\documentclass[usenatbib,onecolumn]{mn2e}
\usepackage{graphicx}
\usepackage{amssymb}

\usepackage{subfigure}
\newcommand{\beq}{\begin{equation}}
\newcommand{\eeq}{\end{equation}}

\def\bar{\overline}

\def\emfb{\overline{\mbox{\boldmath ${\cal E}$}} {}}
\def\bbE{\bar {\bf E}}

\def\beq{\begin{equation}}
\def\ee{\end{equation}}
\def\lsim{\mathrel{\rlap{\lower4pt\hbox{\hskip1pt$\sim$}}
    \raise1pt\hbox{$<$}}}
\def\gsim{\mathrel{\rlap{\lower4pt\hbox{\hskip1pt$\sim$}}
    \raise1pt\hbox{$>$}}}
\def\bfE{{\bf E}}
\def\bfJ{{\bf J}}

\def\bfA{{\bf A}}
\def\bfa{{\bf a}}
\def\bfe{{\bf e}}
\def\bfB{{\bf B}}
\def\bbJ{\bar {\bf J}}
\def\bbV{\bar {\bf V}}

\def\bB{\overline B}
\def\ts{\times}
\def\lb{\langle}
\def\rb{\rangle}
\def\curl{\nabla {\ts}}

\def\bbV{\bar {\bf V}}
\def\bfv{{\bf v}}

\def\bfV{{\bf V}}
\def\bfj{{\bf j}}

\def\bfe{{\bf e}}
\def\bfw{{\bomega}}

\def\bfb{{\bf b}}

\def\bfB{{\bf B}}

\def\bbB{\overline {\bf B}}
\def\bbA{\overline {\bf A}}

\def\div{\nabla\cdot}

\title[Resilience of  helical magnetic fields to turbulent diffusion]
 {On the resilience of  helical 
  magnetic fields to turbulent diffusion and the  astrophysical  implications}
\author [Eric G. Blackman and Kandaswamy Subramanian]{Eric G. Blackman$^{1}$\thanks{E-mail: blackman@pas.rochester.edu} and Kandaswamy Subramanian$^{2}$\thanks{E-mail: kandu@iucaa.ernet.in }\\ $^{1}$Department of Physics and Astronomy, University of Rochester, Rochester NY, 14618, USA\\ $^{2}$IUCAA, Post Bag 4, Ganeshkhind, Pune 411 007, India}
\begin{document}
\date{}
\pagerange{\pageref{firstpage}--\pageref{lastpage}} \pubyear{}
\maketitle
\label{firstpage}
\begin{abstract}
The extent to which  large scale magnetic fields are susceptible to turbulent diffusion is  important   for interpreting the need for in situ large scale dynamos in astrophysics and for observationally inferring field strengths compared to  kinetic energy.  By solving  coupled evolution equations for  magnetic energy and magnetic helicity in a system initialized with isotropic turbulence
and an arbitrarily helical large scale field,  we quantify the decay rate  of the latter for a  bounded or periodic system.  The magnetic energy associated with the non-helical large scale   field  decays at least as fast as the kinematically estimated turbulent diffusion rate, but  the decay rate of the 
 helical part depends  on whether the ratio of its magnetic energy  to the turbulent kinetic energy exceeds a critical value given by $M_{1,c} =(k_1/k_2)^2$, 
where $k_1$ and $k_2$ are the wave numbers of the large and forcing scales. Turbulently diffusing helical fields to  small scales while conserving magnetic helicity requires a rapid increase in total magnetic energy.  As such, only when the  helical field is  subcritical can it  so diffuse.   When supercritical, it decays slowly, at a rate determined by microphysical dissipation even in the presence of  macroscopic turbulence.
In effect,   turbulent diffusion of such a large scale helical field produces small scale helicity whose amplification abates further turbulent diffusion. 
 Two curious implications are that: 
 (1) Standard arguments  supporting the need for in situ  large scale  dynamos   based on  the  otherwise rapid turbulent diffusion of large scale fields
require re-thinking since  only the large scale non-helical field is so diffused in a closed system.  Boundary terms could however  provide  potential pathways for rapid change of the large scale helical field.
    (2) Since     $M_{1,c} \ll1$   for $k_1 \ll k_2$,  
the   presence of long-lived ordered large scale helical fields as in extragalactic jets does not guarantee  that the magnetic field dominates the kinetic energy. 
\end{abstract}
\section{Introduction}

Many astrophysical sources including galaxies, stars,  compact objects, and accretion engines
show direct or indirect evidence for large scale magnetic fields \citep{2004astro.ph.11739S}. 
The extent to which these large scale fields survive in the presence of in situ turbulent diffusion  and the conditions that determine their diffusion rates constrains  the mechanism of their  origin.  Do the fields result  from in situ dynamo generation  or  could they have been  the result of flux freezing  from a previous evolutionary phase?

There has been  debate over the extent to which 3-D turbulent diffusion of large scale fields is effective and the role that the small scale fields play in its potential suppression.
The controversy originated in part from 2-D studies \citep{1991ApJ...376L..21C}
which seemed to suggest suppression. However,   magnetic field lines can interchange 
in 3-D.  This distinction is implicit in the fact that the formalism for computing the isotropic turbulent diffusion coefficient of large scale fields   reveals a suppression in 2-D   that is absent in 3-D \citep{1976JFM....77..321P,1996PhPl....3.1853G}.
The turbulent diffusion of large scale magnetic fields has subsequently been  studied in terms of an effective turbulent diffusion coefficient  for the large scale field,  scaling this coefficient in terms of some power of the magnetic Reynolds number $R_M$ \citep{2002ApJ...579..359B,2005PhR...417....1B,2008ApJ...687L..49B}. 

Most work on the diffusion of large scale fields has not distinguished between the diffusion of helical vs. non-helical  large scale fields.
  An exception is  \citet{yousef03},   which found  that that fully helical large scale fields (where "fully helical large scale fields"  defines  the property that their magnetic energy vanishes when their current helicity vanishes)
 decay more slowly than  non-helical large scale fields in numerical  simulations and discuss this in the context of magnetic helicity evolution.  This stimulates a   quantitative analytic study to understand just how helical the field must  be to be resilient to turbulent diffusion.  As we will see, the suppression of helical field diffusion should be interpreted not  as an intrinsic  suppression of the turbulent diffusion coefficient itself, but as the result of  the current helicity correction
\citep{1975JFM....68..769F,1976JFM....77..321P}  
to the electromotive force which competes with turbulent diffusion.    

One important motivation for this study is the potentially dramatic  implications for interpreting the origin of large scale magnetic fields in astrophysical rotators (galaxies, disks, stars) as  discussed herein.
 An additional motivation   is that jets, particular those in active galactic nuclei (AGN) of  parsec scale, exhibit Faraday rotation that is consistent with an ordered large scale helical field \citep{2008ApJ...682..798A,2008MNRAS.384.1003G,2012JPhCS.355a2019G}. This in turn has led some to conclude that the jets are necessarily magnetically dominated \citep{2005MNRAS.360..869L}.  While Poynting flux dominated  models of jets \citep{2001ApJ...561..915L,2003ApJ...596L.159L,2006MNRAS.369.1167L,2006ApJ...652.1451C}  are plausible,  are  observed  large scale helical fields a definitive signature of a magnetically dominated system?

Although each  class of astrophysical source for  which the evolution of large scale fields plays a role  warrants its own focused study, it is fruitful to investigate simplified problems that potentially identify and elucidate basic principles.  In this spirit,  we focus on the specific underlying physics  of how long it takes for a  large scale magnetic  field to diffuse in the presence of non-helically forced  turbulence when the initial large scale field consists of different helical fractions.  
We study  cases for which the initial field strength does not exceed the kinetic energy as the interiors of astrophysical rotators are typically  not  magnetically dominated.

In section 2 we derive  the basic equations to be solved, drawing from previous work
on 21st century dynamo theory and simplifying the set of equations appropriate for the present problem.  In section 3  we solve these equations.  We discuss the astrophysical implications in section 4 and conclude in section 5.

\section{  Mean Field Decay from Turbulent Forcing in a Closed or Periodic System}

\subsection{Derivation of basic equations}

Here we derive a  system of three differential equations needed to study the  decay of large scale magnetic fields of arbitrary helical fraction in a closed or periodic box. These are the equations for  the time evolution of  (i) large scale magnetic helicity,  (ii) small scale magnetic helicity,  and (iii)  large scale magnetic energy. From these, we will also construct an equation for the evolution of the non-helical large scale field.

To derive the large and small scale magnetic helicity evolution equations we follow standard approaches  \citep{2005PhR...417....1B,2007NJPh....9..309B} and start with the electric field 
\beq
\bfE=-\nabla\Phi -{1\over c}
\partial_t\bfA,
\label{1b}
\ee
where $\Phi$ and $\bfA$ are the scalar and vector potentials. 
Taking the average (spatial, temporal, or ensemble), and denoting averaged
values by the overbar, we have  
\beq
\bbE=-\nabla{\overline \Phi} -{1\over c}\partial_t\bbA
\label{2b}
\ee
Subtracting (\ref{2b}) from (\ref{1b}) gives the equation
for the fluctuating electric field
\beq
\bfe=-\nabla\phi -{1\over c}\partial_t\bfa
\label{3b},
\ee
where $\phi$ and $\bfa$ are the fluctuating scalar and vector potentials.
Using 
$\bbB\cdot \partial_t \bbA= \partial_t(\bbA\cdot \bbB) +c\bbE\cdot \bbB -c\nabla \cdot (\bbA\ts \bbE)$,
where the latter two terms result 
from Maxwell's equation $\partial_t \bbB=-c\curl \bbE$,
and the identity 
$\bbA \cdot \curl \bbE = \bbE\cdot\bbB-\nabla \cdot (\bbA \ts \bbE)$, 
we take the dot product of (\ref{2b}) with $\bfB$ to obtain the evolution of the  magnetic helicity density
associated with the mean fields
\beq
\partial_t(\bbA\cdot\bbB)= -2c\bbE\cdot\bbB
-\div ({c\overline\Phi}\ \bbB + c\bbE\ts \bbA).
\label{5b}
\ee
Similarly, by dotting (\ref{3b}) 
with  $\bfb$ the 
evolution of the
mean helicity density associated with  fluctuating fields is
\beq
\partial_t\overline{\bfa\cdot\bfb}= -2c\overline{\bfe\cdot\bfb}
-\div(c\overline{{\phi} \bfb} + c\overline{\bfe\ts \bfa}).
\label{6b}
\ee

To  eliminate the electric fields 
from (\ref{5b}) and (\ref{6b}) we use Ohm's law with  a resistive term to obtain
 \beq
{\bfE}=-\bfV\ts\bfB/c +\eta \bfJ,
\label{7b}
\ee
where $\bfJ={c\over 4\pi}\curl \bfB$ is the current density and $\eta $ is the resistivity. 
 Taking the average gives 
\beq
{\bbE}=-\emfb/c -\bbV\ts\bbB/c+\eta \bbJ,
\label{8b}
\ee
where $\emfb\equiv \overline{\bfv\ts\bfb}$ is the turbulent electromotive
force.
Subtracting (\ref{8b}) from (\ref{7b}) gives  
\beq
{\bfe}=(\emfb-\bfv\ts\bfb -\bfv\ts\bbB -\bbV\ts\bfb)/c 
+\eta \bfj.
\label{9b}
\ee

Plugging  (\ref{8b}) into (\ref{5b}) and (\ref{9b}) into (\ref{6b}) and  globally averaging (indicated by brackets)  to  ignore divergence terms gives, for the small and large scale contributions respectively
\beq
{1\over 2}\partial_t\lb\overline{\bfa\cdot\bfb}\rb=
-\lb\emfb\cdot\bbB\rb
-\nu_M \lb\overline{ \bfb\cdot\curl\bfb}\rb,
\label{5aa}
\ee
where $\nu_M= (\eta c^2/4\pi)$.
and
\beq
{1\over 2}\partial_t\lb\bbA\cdot\bbB\rb=\lb\emfb\cdot\bbB\rb-\nu_M\lb\bbB\cdot\curl\bbB\rb.
\label{6aa}
\eeq

To obtain an expression for $\emfb$, we use the 'tau'   or   `minimal tau' 
closure approach for incompressible MHD \citep{krr1990,2002PhRvL..89z5007B,2005PhR...417....1B}. This  means replacing  triple correlations by a damping term on the grounds that    the EMF $\emfb$ should decay in the absence of $\bB$.  This gives 
\beq
\partial_t \emfb = \lb \partial_t \bfv  \times \bfb \rb + \lb \bfv \times \partial_t \bfb \rb 
={\alpha \over {\tilde \tau}}\bbB-{\beta\over {\tilde \tau}}\curl\bbB-\emfb/{\tilde \tau},
\label{7aa}
\ee
where ${\tilde \tau}$ is a damping time 
and 
\[
\alpha \equiv {{\tilde \tau}\over 3} \left({\lb \bfb\cdot\curl\bfb \rb\over 4\pi\rho}- 
\lb \bfv\cdot\curl\bfv \rb\right) \quad  {\rm and} \quad 
\beta\equiv   {{\tilde \tau} \over 3}\lb v^2\rb.
\]

The time evolution of  $\emfb$ can be retained as a separate equation to couple
into the theory and solve, but  simulations of magnetic field evolution in
forced isotropic helical turbulence 
 reveal that a good match to the large scale magnetic field evolution
in  simulations can be achieved even when the left side of (\ref{7aa}) is ignored  and 
 ${\tilde \tau}={1\over v_2k_f}$, the eddy turnover time associated with the forcing scale
 \citep{2002ApJ...572..685F,2005A&A...439..835B}. 
We adopt that  approximation here.  Rearranging  (\ref{7aa}) then gives
\beq
\emfb =\alpha \bbB-\beta \curl\bbB,
\label{emfb}
\ee
Eqs. (\ref{5b}) and (\ref{6b}) then become
\beq
{1\over 2}\partial_t\lb\overline{\bfa\cdot\bfb}\rb=
-\alpha\lb\bB^2\rb +\beta\lb\bbB\cdot\curl\bbB\rb
-\nu_M \lb\overline{ \bfb\cdot\curl\bfb}\rb
\label{5ab}
\ee
and
\beq
{1\over 2}\partial_t\lb\bbA\cdot\bbB\rb=
\alpha\lb\bB^2\rb -\beta\lb\bbB\cdot\curl\bbB\rb
- \nu_M\lb\bbB\cdot\curl\bbB\rb.
\label{6ab}
\ee

Note that the  energy associated with the small scale magnetic field does not enter  $\emfb$ above.
 Therefore it does not  couple into equations (\ref{5aa}) and (\ref{6aa}).  
 It appears  only as a  higher order hyperdiffusion correction \citep{2003PhRvL..90x5003S}  which  we neglect because its ratio to the $\beta$ term in the EMF is  ${ b^2\over 4\pi \rho v^2}{k_1^2\over k_2^2}<<1$ .
However, upon plugging (\ref{emfb}) into 
those equations, the energy associated with the large scale field $\bB^2$ {\it does} enter.
Therefore we need a separate 
equation
for the energy associated with the energy of the mean field.
To obtain this equation we dot $\partial_t \bbB = -c\curl \bbE$
with $\bbB$ and ignore the flux terms to obtain
\beq 
\begin{array}{r}
{1\over 2}\partial_t \lb \bB^2\rb =-c\lb \bbB\cdot \curl \bbE\rb=
-c\lb \bbE\cdot \curl\bbB\rb
=\lb\emfb\cdot \curl\bbB \rb
-\nu_M\lb (\curl\bbB)^2 \rb
=
\alpha \lb \bbB\cdot\curl \bbB\rb
 - \beta \lb (\curl\bbB)^2 \rb
-\nu_M\lb (\curl\bbB)^2 \rb,
\end{array}
\label{ind}
\ee
where the latter two similarities follow from using
(\ref{8b}) and (\ref{emfb}) and 
$\bbV=0$.

Eqs.  (\ref{5ab}), (\ref{6ab}), and (\ref{ind})  form a  set  that can be solved in a two scale model as long as $\lb v^2\rb$ is supplied by steady forcing such that $\partial_t\lb v^2\rb\simeq 0$, 
and  $\lb\bfv \cdot\curl \bfv\rb$
remains small.  The  implications and  justification  of this latter assumption for present purposes
will be discussed in  more detail  after the results of solving the above equations are presented.

\subsection{Two Scale Model and Dimensionless Equations}

To extract the essential implications of the coupled  Eqs. (\ref{5ab}), (\ref{6ab}), and (\ref{ind})  for a closed or periodic system, we adopt a standard two-scale model \citep{2002PhRvL..89z5007B,2005PhR...417....1B} 
 and indicate  large scale mean
quantities with  subscript "1" and fluctuating quantities with   subscript "2." 

We write the wave number  $k_1>0$   to be that associated with the variation of large scale quantities and that the wave number $k_2>>k_1$ to be that associated with the variation of small  scale  quantities.      We also assume that  $k_2=k_f$ where $k_f$ is the forcing wave number  at which 
$v_2^2 = \lb v^2\rb$ is maintained to be a constant.  
We also assume the  turbulence is non-helically forced   (i.e. initially driven with  $(\lb\bfv\cdot\curl \bfv \rb)= 0$ and  subsequently  $|\lb\bfv\cdot\curl \bfv\rb|<<|\lb\bfb\cdot\curl \bfb\rb|/(4\pi\rho)$,
 an assumption to be discussed further in section 3.5.

Applying this two scale approximation  to a closed or periodic system,  we   then freely use
 $\lb \bbB \cdot \curl \bbB \rb = k_1^2 \lb \bbA \cdot \bbB \rb$, and
    $\lb (\nabla \times \bbB)^2\rb = k_1^2\lb \bbB^2\rb$, along with   
    $\lb \bfb\cdot \curl \bfb \rb = k_2^2 \lb \bfa \cdot \bfb \rb$.
 Eqs (\ref{5ab}), (\ref{6ab}), and (\ref{ind})  then become
\beq
\partial_t H_1= \left({2{\tilde \tau}\over 3}\right)k_2^2 H_2{B_1^2\over 4\pi \rho} - {2{\tilde \tau}\over 3} v_2^2k_1^2 H_1-2\nu_M k_1^2 H_1,
\label{6}
\eeq
\beq
\partial_t H_2= -\left({2{\tilde \tau}\over 3}\right)k_2^2 H_2{B_1^2\over 4\pi \rho} + {2{\tilde \tau}\over 3} v_2^2k_1^2 H_1-2\nu_M k_2^2 H_2,
\label{7}
\eeq
and
\beq
\partial_t B_1^2 = \left({2{\tilde \tau}\over 3}\right)\left({k_2^2 H_2 k_1^2 H_1 \over 4\pi \rho}\right) - {2{\tilde \tau}\over 3} v_2^2 k_1^2 B_1^2-2\nu_M k_1^2 B_1^2,
\label{8}
\eeq
where $H_1= \lb\bbA\cdot\bbB\rb$ and 
$H_2= \lb\overline{\bfa\cdot\bfb}\rb$,
$B_1^2 = \lb\bbB^2\rb$.

We  non-dimensionlise these  equations by scaling lengths in units of $k_2^{-1}$,
and time in units of $\tau=(k_2 v_2)^{-1}={\tilde \tau}$, where the latter equality follows since $k_f=k_2$ in our two-scale approach. 
We define
\[h_1\equiv {k_2 H_1\over 4\pi \rho v_2^2}, \
h_2\equiv {k_2 H_2\over 4\pi \rho v_2^2}, \
R_M\equiv {v_2\over \nu_M k_2}, \ 
{\rm and}\  
M_1\equiv{\lb\bbB^2\rb\over 4\pi \rho v_2^2}.\]
Eqs. (\ref{6}), (\ref{7}), and (\ref{8}) can then  be respectively written as
\beq 
\partial_\tau h_1 = {2\over 3}h_2 M_1-{2\over 3}\left({k_1\over k_2}\right)^2 h_1-{2\over R_M}\left({k_1\over k_2}\right)^2h_1,
\label{9}
\eeq
\beq
\partial_\tau h_2 = {-2 \over 3}h_2 M_1+{2\over 3}\left({k_1\over k_2}\right)^2 h_1-\left({2\over R_M}\right)h_2,
\label{10}
\eeq
and
\beq
\partial_\tau M_1 ={2 \over 3} h_1 h_2\left({k_1\over k_2}\right)^2 - {2 \over 3}M_1 \left({k_1\over k_2}\right)^2
 - \left({2\over R_M}\right) M_1\left({k_1\over k_2}\right)^2.
\label{11}
\eeq

We can define a fully helical large scale field by the property  ${c\over 4\pi}
\vert\lb\bbJ\cdot\bbB\rb\vert /k_1 = k_1\vert\lb\bbA\cdot\bbB\rb\vert = \lb\bbB^2\rb$.  If we choose  a positive large scale helicity $h_1>0$,  we can drop the absolute value and 
 divide the large scale magnetic energy into a  fraction proportional to the magnetic  (or current) helicity 
$f _1\equiv  (k_1\lb\bbA\cdot\bbB\rb)/(\lb\bbB^2\rb)$ and a fraction independent of the magnetic helicity  $(1-f_1)$.  Multiplying (\ref{9})  by $k_1/k_2$ and subtracting it from  (\ref{11}), the evolution equation for the non-helical contribution to the large scale magnetic energy  $M_{1,nh}\equiv M_1-k_1 h_1/k_2$ becomes
\beq
\partial_\tau M_{1,nh} =-{2\over 3} M_{1,nh}\left({k_1\over k_2}h_2  + { k_1^2\over k_2^2}\right)
 - \left({2\over R_M}\right) M_{1,nh}\left({k_1\over k_2}\right)^2,
\label{11aa}
\ee
which has all decay terms and implies a decay rate even faster than 
that given by the turbulent diffusivity alone when $h_1,h_2>0$. 
This will be important in understanding the  solution 
plots that follow in the next section.

Note  that our definition of " helical large scale" field via the above decomposition of the magnetic energy makes use only of quadratic functions of the large scale field.
  We do not require any meaning beyond this decomposition for present  purposes. Note also that the large scale field represents an averaged part of the physical field, not the full physical field.
  Thus the topology of the large scale field can be different from  the  topology of the full field.
\section{Discussion of Solutions}

Here
we discuss the solutions  of Eqs. (\ref{9}), (\ref{10}), and (\ref{11}) 
for  several different cases, assuming that 
the kinetic energy  per unit mass is kept steady, and driven by non-helical 
forcing at  $k_f=k_2$.
We identify and derive a  minimum helical magnetic energy, in units of kinetic energy, required  for slow decay.

\subsection{Solutions for fixed initial magnetic energy but varying initial magnetic helicity fraction}

Solutions to Eqs. (\ref{9}), (\ref{10}), and (\ref{11}) are shown in Figs. 1a-d for $R_M=800$
and $k_1=1$ and $k_2 = 5$.
 Each  curve in each panel represents a solution with a different initial helical fraction  $f_{1,0}$
of  large scale magnetic energy, with the initial large scale magnetic energy set to equipartition with the kinetic energy, i.e. $M_{1,0}= 1$.  All cases start with  $h_2(t=0)=0$.   The six curves of  progressively increasing dash spacing  correspond  to   $f_{1,0}=0.95, 0.7, 0.5,0.2, 0.04,0.004$ respectively, so that the solid lined curves correspond to $f_{1,0}=0.95$ and the  widest spaced dashed curves  correspond to $f_{1,0}=0.004$.   Fig.1a shows the  time evolution of the large scale helical magnetic energy  $M_{1,h}={k_1\over k_2}h_1$ divided by  ${k_1h_c\over k_2}$, where $h_c=k_1/k_2$ is the  critical magnetic helicity  derived  in Sec. 3.3 below.    The large scale non-helical magnetic field energy  $M_{1,nh}$ (Fig 1b), and the  total large scale magnetic energy (Fig 1c)  $M_1$ are normalized to the initial non-helical magnetic energy $M_{1,nh0}=M_{1,0}-{k_1\over k_2}h_1$ and the initial total magnetic energy $M_{1,0}$  respectively.  The evolution of the non-helical magnetic energy follows Eq. (\ref{11aa}), derived  
   by subtracting   $k_1/k_2$  times  (\ref{9}) from  (\ref{11}).  Fig. 1d shows the time evolution of $h_2/h_c$.
 
All curves  of Fig 1b show that  the non-helical field decays rapidly for all values of $f_{1,0}$. The  minimum decay rates occur for  $h_2=0$. When $h_2 >0$, the rate of decay is even faster than turbulent diffusion,  as can be seen from equation (\ref{11aa}) in which the first term on the right side 
 provides enhanced decay for $h_2 >0$.  The rapid decay of the non-helical field in all cases also implies that during the slow decay regimes of  $M_1$ in Fig. 1c, the total field is primarily helical.  

The slow decay regimes in  Figs 1a and 1c correspond to decay  at a   microphysical dissipation rate,  determined by the last term in Eq. (\ref{9})).   As the plots show,  in  these regimes   $h_1(t) /h_{1,c}> 1$.  When  instead $h_1(t) /h_{1,c}< 1$,  the decay is  much more rapid and   determined by  turbulent diffusion---the penultimate term of (\ref{9}).  Correspondingly, when the helical field is  subcritical right from the start,  the helical field rapidly decays. 
 
 \subsection{Solutions for fixed initial magnetic helicity fraction but varying  initial magnetic energy and scale ratio}

For the solutions to Eqs. (\ref{9}), (\ref{10}), and (\ref{11}) 
shown in  Fig. 2,  we used   $f_{1,0}=0.999$ for all curves and varied the initial
 magnetic energy $M_{1,0}$.  We  used $k_1=1$ and $k_2=5$ with $R_M=800$.
 From top to  bottom the curves in Fig 2a  correspond to  
${M_{1,0}\over (k_1/k_2)^2} =20, 5,1,0.5,0.1, 0.01$, respectively.
The third curve from the top 
corresponds to our analytically derived critical value  (see next subsection) $M_{1,0}=k_1 h_c/k_2=(k_1/k_2)^2$.
This curve marks the approximate demarcation line between slow and fast decay  curves.
For curves with  $M_{1,0} \gg (k_1/k_2)^2$ decay is slow (resistive), whereas for $M_{1,0} \ll (k_1/k_2)^2$  decay is fast (unfettered turbulent diffusion).

In Fig. 3 we show solutions to Eqs. (\ref{9}), (\ref{10}), and (\ref{11}) 
for  $f_{1,0} =0.999$ but  for different values of $k_1/k_2$ and $R_M$ than the values used  in Fig 2. 
For Fig. 3a,   $k_1=1$, $k_f = 10$,  $R_M = 8000$, and  for Fig. 3b,   $k_1=1$, $k_f = 20$,  $R_M = 8000$, and $f_{1,0} =0.999$.    In each panel, the curves from top to bottom 
correspond to  initial energies ${M_{1,0}\over (k_1/k_2)^2 }=5,2,1,0.5,0.1, 0.01$ respectively.
The  third curve from the bottom in each panel  again corresponds to our analytically estimated critical  
curve bounding the slow and fast decay regimes.  All curves decay more gradually  in Fig. 3b  than Fig. 3a.  (note the difference in scale of the $y$-axis in the two figures) because  both the turbulent  
diffusion  and the resistive diffusion terms  for the case considered in Fig. 3b,
are correspondingly  reduced by the smaller value of  $(k_1/k_2)^2$.

\subsection{Derivation of the critical helicity $h_{c}$ }

We now derive the critical helicity $h_c$ and the associated critical helical magnetic energy by 
noting that the   slow decay regime requires   the last term on the right of (\ref{9}) to be 
at least comparable to the sum of the  first two terms on the right.  For large $R_M$, each of those first  two terms is separately much larger than the last term.   Therefore the first two terms must approximately balance.  These same terms also appear in the equation for $h_2$, implying that it too evolves slowly (as  seen in Fig. 1d.) in the slow decay regime.   Since initially we always consider  $h_1>0$ and $h_2=0$,  emergence to a slow decay regime implies  a rapid 
evolutionary phase (with negligible dependence on  $R_M$) where the buildup of $h_2$
leads to an approximate balance between these two terms.  But if there is not enough 
$h_1$, then there is not enough supply of magnetic helicity to grow the required $h_2$ to establish the slow decay regime.

We can estimate  the needed amount of $h_2$ that must be grow by  balancing first two terms on the right of  either   (\ref{9}) (or (\ref{10})) to obtain
that 
\beq
 h_2 M_1 \simeq  \sim (k_1/k_2)^2  h_1.
\label{11c}
\eeq
Since $M_1\simeq k_1 h_1 /k_2$ in the slow decay regime,
Eq. (\ref{11c})  gives the result that $h_2 \simeq  k_1/k_2$ in this regime.
The only possible source of $h_2$  is $h_1$ given our initial conditions, 
therefore the above value of $h_2$  gives the minimum required of $h_1$  to achieve the slow decay decay regime.  That is, we must have $h_1>h_c\equiv  k_1/k_2$ for a slow decay regime of the large scale helical field. If ever $h_1 << h_{c}$,  the large scale field will decay rapidly.
Identification of this  critical helicity $h_c$ is the key to  explaining the curves  shown in Figs. 1,2, and 3.

The critical magnetic energy associated with  $h_c$ is simply
$M_{1,c} = k_1 h_{c}/k_2=(k_1/k_2)^2$ and  this can be substantially below equipartition when $k_1/k_2 \ll 1$.  The amount of magnetic energy that decays slowly, $M_{1,slow}$,  is then the difference between the magnetic energy contained in the helical field $M_{1,h}=f_1 M_1=k_1h_1/k_2$ and  $M_{1,c}$.  Dividing this difference by the total magnetic energy then gives the fraction of energy that will decay slowly, namely
\beq
{M_{1,slow}\over M_1} =Max \left[{M_{1,h}-M_{1,c}\over M_1},0\right]=
Max \left[f_1- {k_1^2\over k_2^2}{1\over  M_1},0\right].
\label{18}
\eeq
Eq. (\ref{18}) shows that  most of the initial magnetic energy can decay slowly 
even if the system is not magnetically dominated or maximally helical. For example, we used $M_{1,0}=1$ (equipartition between total magnetic and kinetic energy) and $k_1/k_f= 1/5$ for the solution of Fig.  1 so that  a fraction   ${M_{1,slow}\over M_1}=f_1 - {1\over 25}$ of the initial magnetic decays slowly for $f_1\ge 1/25$.

\subsection{Resilience of helical fields to diffusion is not a reduction in the diffusion coefficient}

The resilience of the helical field to turbulent diffusion as shown above, 
is  the result of  the current helicity part of the $\alpha$ effect in the language of dynamo
theory, not an intrinsic change in  the diffusion coefficient $\beta$. 
This is an important distinction because the  $\beta$ term 
also appears in the non-helical magnetic energy evolution equation  where it is not abated
by terms involving magnetic helicity,  but  instead can even be enhanced by them (Eq. \ref{11aa}).
The point is that the helical and non-helical large scale fields obey different equations.
Parametrization 
of $\beta$ in terms of $R_M$ (e.g. \citet{2002ApJ...579..359B,yousef03}) can be misleading  in this respect, because $\beta$ itself does not change even when the helical field decays at the resistively limited rate.

\subsection{Neglect of  kinetic helicity evolution for our choice of  initial conditions}

 In general the $\alpha$ effect is the difference between  small scale current and kinetic helicity,
 but we have not included an equation for the time evolution of the kinetic helicity.
If the growth rates of these two helicities  were equal so as to keep $\alpha\sim 0$, then 
the large scale helical field would decay as fast as the non-helical field. 
We now discuss our  justification for ignoring kinetic helicity evolution for our specific choice of initial conditions.
 
  \citet{2004PhPl...11.3264B} and  \citet{2012MNRAS.423.2120P}  
showed that the small scale kinetic helicity can grow significantly when  the system is  initiated with
fully helical small scale fields. But  the small scale current helicity can only drive small scale kinetic helicity  growth  and conserve magnetic helicity by  inverse cascading,  bringing helical magnetic energy  up to  larger scales.   In our present case, any initial magnetic helicity is solely on the large scale to begin with   and  we now argue that the kinetic helicity  associated with the small scale is not expected to grow significantly.

The kinetic helicity growth in the two-scale approximation for a closed system is given by \citep{2004PhPl...11.3264B}
\beq
{1\over 2}  \partial_t H_2^V=\left(\bbB\over4\pi  \rho\right)\cdot\lb{\bfw }\ts\bfb\rb(k_2-k_1)
-\nu k_2^2\lb\bfv \cdot \bfw\rb
+{1\over c\rho}
\lb {\bfw}\cdot (\bfj\times\bfb)\rb+{1\over c\rho} \lb \bfv \cdot \curl (\bfj\times\bfb)\rb,
\label{h2v}
\ee
where $\nu$ is the viscosity and 
we assume any non-helical forcing function  in the velocity equation does not explicitly contribute.   
We expect that the first term on the right can only grow kinetic helicity if the contribution from
either
$\bfv$ or $\bfj$ to the correlation comes from their helical part.  So we can look
at this term in two ways, either focusing on the velocity contribution or the  magnetic field contribution.
We consider the latter approach first. If we ignore  total  divergence (surface) terms
 and spatial derivatives of the cross helicity $\lb\bfv\cdot\bfb\rb$ (given that the latter 
 evolves  only via decay  $\partial_t\lb\bfv\cdot\bfb\rb = -k_2^2(\nu+\nu_M)\lb\bfv\cdot\bfb\rb$), 
 then $\lb{\bfw}\ts \bfb \rb_q 
= \lb v_s \partial_q b_s\rb=\lb \bfv \ts \bfj\rb_q$
and the first term on the right of (\ref{h2v}) can be written  $2\bbB\cdot\lb{\bfv}\ts\bfj\rb(k_2-k_1)$. 
For a maximally helical small scale field 
 $\bfj\times \bfb=0$   (though $\bbJ\times\bfb \ne 0 \ne \bfj \times \bbB$)  
and so
the last two terms of (\ref{h2v}) vanish. If we   consider the case $\bfj\cdot\bfb>0$, 
then $2\bbB\cdot\lb{\bfv}\ts\bfb\rb k_2(k_2-k_1) = 2\emfb\cdot\bbB k_2(k_2-k_1)$ 
and $H_2^V$ could grow positive as quickly as the current helicity 
$k_2^2 H_2$ since the latter grows at a rate determined by multiplying 
  the first term on the right of  Eqn. (\ref{5aa}) by $k_2^2$. 
  
 But since we expect only the helical fraction of the small scale magnetic field to 
 grow kinetic helicity, for non-maximally helical small fields, the  factor  $\lb\bfv\times\bfj\rb$ would  be reduced to
$\sim f_2 k_2\lb\bfv\times\bfb\rb$ where $f_2\equiv \left({\lb\bfj\cdot\bfb\rb^2 \over\bfj^2\bfb^2}\right)^{1/2}\le1$ is the helical magnetic field fraction at  the small  scale. 
The solutions shown in Fig. 1 indicate that  $k_2 H_2 \le k_1v^2/k_2$ even
when  no drain  into $H_2^V$ is considered. Since we would   expect the  non-helical turbulent forcing  to produce $\lb b^2\rb\sim\lb v^2\rb $, we would  then have     $f_2 =| k_2 H_2|/\lb b^2 \rb \le k_1/ k_2$ 
for all 
$f_1\le 1$,  and the growth term of $H_2^V$  on the 
right of (\ref{h2v}) would  be less than $ k_1/k_2$  times that  on the right of (\ref{5aa}). 
In addition,  when $\bfj \times \bfb\ne 0$,
the  triple correlation terms on the right of (\ref{h2v})
 survive. 
 Since  these  depend only on the non-helical field,  we expect them to be  
decay  terms.  
In addition, an aspect of the kinetic helicity evolution that is not well captured in a two-scale theory is that for typical inertial range spectral power laws,  the microphysical viscous diffusion of kinetic helicity diverges  with increasing Reynolds number, unlike that of magnetic helicity  (e.g. \citet{2005PhR...417....1B}).  This exacerbates the relative importance of microphysical diffusion in the kinetic helicity equation compared to that in the magnetic helicity equation.

Now consider the second possibility that
the first term on the right of (\ref{h2v}) 
survives
only when there is a helical velocity field with arbitrary magnetic field.
The magnitude of the right  of (\ref{h2v}) can then  be written $ 2  f_v  \emfb\cdot \bbB  k_2(k_2-k_1)$,  roughly a factor of the fractional kinetic helicity $f_v \equiv {H_v  \over | \omega| |v| }$
slower than  time evolution of  the current helicity, $k_2^2 H_2$. (The latter again evolving at a rate determined by multiplying   the first term on the right of Eqn. (\ref{5aa}) by $k_2^2$.) 
Since the kinetic energy is forced  non-helically, $f_v << 1$ initially and it is likely that
$\emfb\cdot\bbB$ would already saturate before significant $H_2^V$ could grow.
 
 For the above reasons, we therefore expect  that $H_2^V$ would not grow significantly  to affect our solutions 
 to the specific initial value problem  presented in the previous subsection.

\section{Astrophysical Implications}

Our results show that, when subjected to steady non-helical turbulent forcing of kinetic energy at wavenumber $k_2$,  the non-helical large scale field decays at least as fast as the unfettered turbulent diffusion rate, but the large scale helical field rapidly decays only when  the ratio of its  energy 
to the turbulent kinetic energy drops below the critical value $M_{1,c}=(k_1/k_2)^2$.
 Above this value, the helical field  decays on a microphysical resistive time scale.  With the caveat that we have not included  boundary terms or buoyancy,  this has several provocative implications.

\subsection{Presence of observed large scale helical fields does not guarantee a magnetically dominated plasma}
Observations of large scale  helical fields, such as those detected by  Faraday rotation in extragalactic jets \citep{2008ApJ...682..798A,2012JPhCS.355a2019G},
or inferred in gamma-ray bursts \citep{yonetuko12},  are sometimes interpreted to 
imply that the field is force-free and therefore dominates the kinetic energy of the system.   Our calculations show that this is  not necessarily the case: The fact  that $M_{1.c} \ll 1$ for $k_1\ll k_2$, 
implies that  even significantly  sub-equipartition helical fields  decay on resistive time scales 
which are typically much longer than dynamical jet time scales.  If a jet contained  isotropic or  quasi-isotropic MHD turbulence,   perhaps supplied via an instability at the radial interface between jet and ambient medium \citep{2008A&A...488..795R},  then the large scale helical field could survive intact and the system would not necessarily be magnetically dominated.  
Although the helical large scale field could appear force-free in the sense  that $\bbJ\times\bbB\simeq 0$,  this does not mean that $\bfj\times \bbB$ or $\bbJ\times \bfb$ vanish. The non-vanishing of the latter are essential in the  derivation of (\ref{6}) and (\ref{7}), and particularly the 
appearance of $H_2$ (the driving due to the current helicity) 
on the right  sides.

We did not include any anisotropic velocity such as shear  in our calculations. Nevertheless, our results  still demonstrate that the basic point that  mere observation of a helical large scale field does not {\it prove} magnetic energy dominance.

\subsection{Rethinking when  in situ dynamos are needed to produce large scale fields}

Taken at face value, the survival of   helical fields to turbulent diffusion
  may reduce the essentiality of in situ dynamos in systems if boundary terms are unimportant.
  
Consider  the case of galaxies: A long standing criticism of relying on  primordial or protogalactic fields as the primary source of galactic fields has been that the mean field would otherwise rapidly  diffuse in the galaxy via supernovae induced turbulence 
if this turbulence were unable to also facilitate competitive exponential growth from  a large scale dynamo   \citep{2004astro.ph.11739S}.   Our calculations  provide  rejuvenated credence to  pre-galactic mechanisms of  large scale field production
\citep{
2008RPPh...71d6901K,
2010AN....331..110S,2012SSRv..166...37W} ,  
 and specifically those that produce  sufficiently strong  helical fields 
\citep{2000PhRvD..62j3008F,2008PhRvL.101q1302C,2008PhRvL.100x1301D,2012JCAP...06..008S} 
because only such helical fields would avoid diffusing  over a galactic lifetime in  the absence of boundary terms.

Most of the energy in large scale galactic magnetic fields resides in non-helical toroidal fields  amplified from poloidal fields by differential rotation. As long as the turbulent decay time for the non-helical field exceeds the linear shear time, then we can expect a predominance of non-helical field in a steady state, even without an in situ dynamo to regenerate the poloidal fields. This is because  the helical field provides a minimum value below which the toroidal field cannot drop. The toroidal field  enhancement over the poloidal field would be  that which can be linearly amplified in a non-helical field diffusion time.    

One distinguishing signature  of a primordial helical field would be that its magnetic helicity would be 
 of one sign and would thus not show a reversal across the mid-plane of a rotator.  In contrast, an in situ large scale dynamo would be expected to produce field whose helicity changes sign across the mid-plane  because of  the reflection asymmetry of  transport coefficients that drive the field growth.  
This leads to predictions for relative signs of  large and small scale helicity and their respective
sign reversals for the sun \citep{bb03} which seem to be observed \citep{2011ApJ...734....9B},  thus providing  evidence for in situ large scale dynamo  action in the sun. We lack such measurements of galactic fields, but the  absence of a large scale magnetic helicity reversal across the mid-plane would be evidence for primordial galactic fields.

As mentioned, an important caveat is that our calculations do not include  buoyancy or other boundary loss terms that could extract  large scale helicity at a rate that may still need to be re-supplied  from within  the rotator.  If such terms are important, then both helical and non-helical large scale fields would deplete and an  in situ dynamo would be needed for replenishment.  But this shifts the focus from turbulent diffusion in conventional wisdom to that of boundary loss terms in assessing the   necessity of in situ dynamos. In fact, it may be the boundary loss terms that also facilitate
such dynamos in the first place by ejecting small scale helicity that would otherwise 
quench the large scale dynamo  \citep{2000ApJ...534..984B,2001ApJ...550..752V,2006A&A...448L..33S,2007MNRAS.377..874S}. 

Similar considerations regarding the survival of helical fields  would apply  for the large scale fields of stars and accretion disks. A long standing debate over whether the large scale fields that power jets must be produced in situ or survive advection in turbulent disks to grow by flux freezing has persisted \citep{1994MNRAS.267..235L,2009ApJ...701..885L}. 
Our results here would suggest that any helical part of these large scale fields would be more resilient to diffusion and more easily advected.
Again  boundary loss terms could change the story in that such terms would both justify the need for a dynamo when it comes to ejection of large scale helical  fields, while also potentially being essential to its operation by the ejection of small scale magnetic helicity.

Our  results would also suggest that, in the absence of boundary terms,   fast cycle periods in stars and disks could only depend on a rapid diffusion of  non-helical fields, since the helical fields 
diffuse too slowly.    It should be noted that even in a closed box, when shear is included,  cycle periods can arise with $\lb\bbB^2\rb$ and $\lb \bbA\cdot\bbB\rb$ remaining constant \citep{2002ApJ...579..359B}. The extent to which such closed volume cycle periods can be  fast or remain resistively slow needs more study.

\section{Conclusion}

Using an incompressible two-scale theory that couples magnetic helicity evolution and large scale magnetic energy evolution, we  have quantified the relative decay rates of helical
and non-helical large scale magnetic fields subject to isotropic non-helical turbulent forcing in a closed (or periodic) system. We identified a critical ratio of the large scale helical magnetic energy to the turbulent kinetic energy given by  $M_{1,c} =(k_1/k_2)^2 $ above which  helical magnetic energy is 
immune to turbulent diffusion and decays at the microphysical resistive rate.   In contrast, we find that the non-helical field always decays at least as fast as the turbulent diffusion rate and approaches a factor of two faster  when the small scale helical field  reaches its maximum.  
The calculations herein corroborate  the slow decay of helical fields  seen in  \citep{yousef03} and provide a more complete theoretical explanation by quantifying how much helical field can survive turbulent diffusion and how much decays.

The physical interpretation of our result emerges from  basic principles of magnetic helicity dynamics.  For a fixed amount  of total magnetic helicity, magnetic energy is minimized when magnetic helicity resides on the largest scale.  This is the state toward which the system would relax in the absence of kinetic forcing \citep{1986RvMP...58..741T}. 
If the system is to conserve magnetic helicity and the large scale helical field is to  turbulently diffuse, then the small scale helical field  must gain energy beyond than that contained in the initial large scale field.  The source of the extra energy is the non-helical kinetic forcing.  However any growth of small scale magnetic helicity supplies current helicity to the dynamo $\alpha$ coefficient, which in turn regrows large scale helical field---the inverse cascade.    A quasi-steady state results when there is enough magnetic helicity in the system such that the inverse cascade from the small scale sends magnetic helicity back to large scales at a rate competitive with of the turbulent diffusion.  This state requires an energy in the small scale helical field equal  to just $k_1/k_2$ of the kinetic energy.
Correspondingly,  the initial large scale helical field energy required to supply the necessary 
magnetic helicity is  the further reduced fraction $(k_1/k_2)^2$ of  the kinetic energy.

All of this leads to a rethinking of the conditions for when in situ dynamos are required in astrophysical objects: If helical fields survive turbulent diffusion, then they would only decay on microphysical resistive time scales, obviating the need for  a source of in situ amplification.
 This  may reinvigorate the potential relevance of primordial helical fields for galaxies
 \citep{2000PhRvD..62j3008F,2008PhRvL.101q1302C,2008PhRvL.100x1301D,2012JCAP...06..008S,
2012SSRv..166...37W} 
and advected helical fields for large scale jets in accretion disks.
\cite{2009ApJ...701..885L}.  The fact that large scale helical fields can survive even when sub-equipartition with the turbulent kinetic energy also highlights that the mere appearance  of a force-free-looking large scale field does not prove that the magnetic energy dominates the kinetic energy.

Our present calculations  ignored kinetic helicity evolution, which we argued to
be a small contributor to the class of initial value  problems studied.  We also ignored  boundary terms and buoyancy, two key ingredients of real systems. Generalizations that include these ingredients are of great interest for future work. If, in the context of astrophysical disks and stars, buoyancy removes large scale fields without discretion as to their helicity, then in situ dynamos would  be needed  to sustain large scale fields.  Work in 21st century  large scale dynamo theory
 has also been evolving toward the perspective that helicity fluxes, if not boundary terms, 
may actually be essential for the operation of  large scale  dynamos by removing the small scale helicity that clogs the evolution of large scale helicity.  Thus boundary loss terms for the large scale field may require a dynamo whereas the dynamo itself might require the boundary loss terms for the small scale field.

In this context,  note that our initial condition invoked a net helical large scale field without a compensating  helical field of opposite sign. This circumstance itself  could arise from an  MHD dynamo only if the compensating magnetic helicity of opposite sign were dissipated or lost via  a boundary flux. 
Had we started with equal and opposite helical fields on small and large scales
subjected to our same non-helical turbulent forcing, the small scale magnetic helicity would inverse cascade and annihilate the large scale magnetic helicity rapidly and the total large scale field would
decay rapidly. The slow diffusion of the large scale helical  field that we have studied depends on having a finite net magnetic helicity as a starting point.

\section*{Acknowledgments} 
We thank  A. Brandenburg, G. Field, E. Quataert, K. Park, F. Nauman and L. Chamandy 
for related discussions.
KS was visiting  the Univ. of  Rochester when this work
began and both EB and KS  acknowledge  support from 
 NSF Grants PHY-0903797 and AST-1109285 during this visit.

\vfill 
\eject
\newpage
\begin{figure}
\centering
\mbox{\subfigure[]{\includegraphics[width=2.5in]{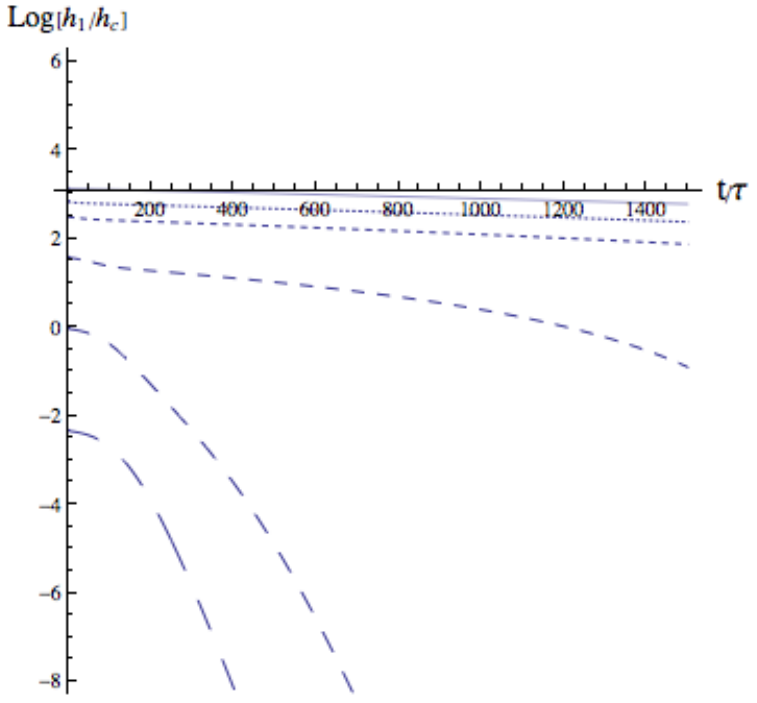}}\quad
\subfigure[]{\includegraphics[width=2.5in]{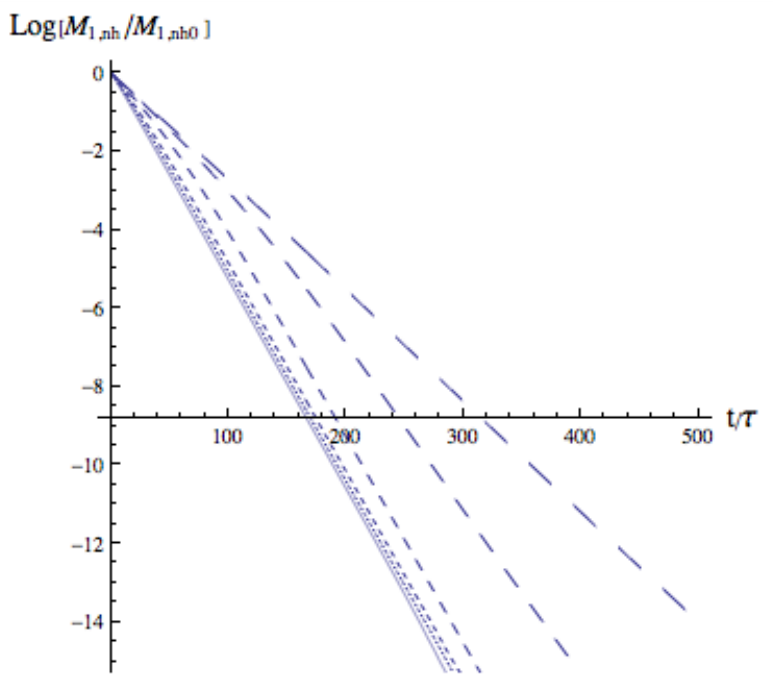}}}
\mbox{\subfigure[]{\includegraphics[width=2.5in]{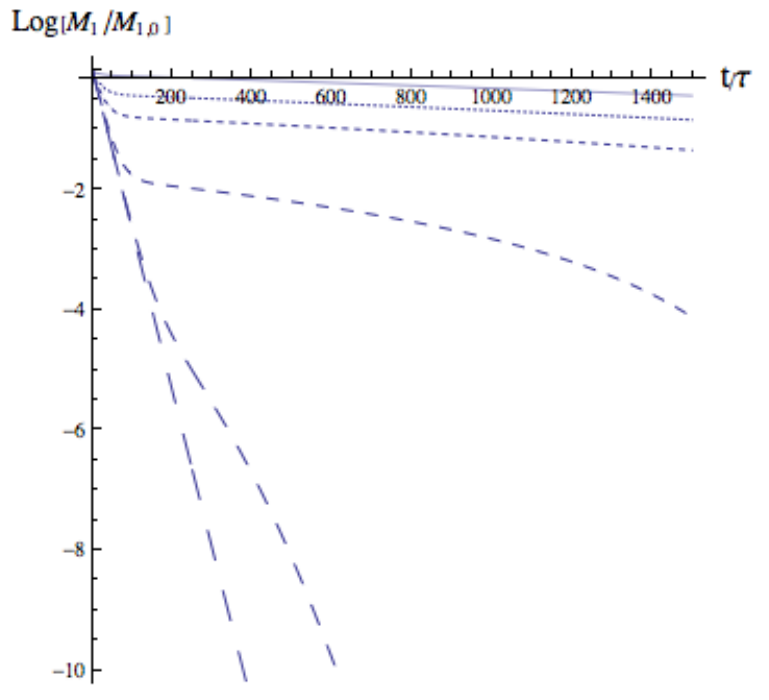}}\quad
\subfigure[]{\includegraphics[width=2.5in]{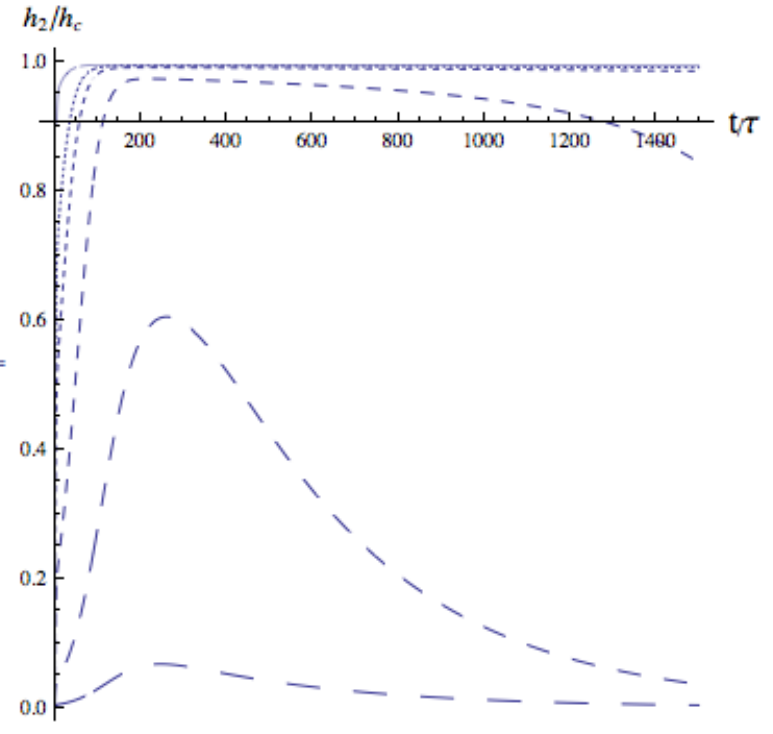}}}
\caption{ Solutions to Eqs. (\ref{9}), (\ref{10}), and (\ref{11}) for $k_1=1$, $k_2=5$, $R_M=800$,
 constant $v_2^2$,
 and  $M_{1,0}= 1$ for all curves.
In each panel. the six curves of successively increased dash spacing correspond  to    $f_{1,0}=0.95, 0.7, 5,0.2, 0.04,0.004$ respectively. (a) Dimensionless large scale magnetic helicity, where $h_c=k_1/k_2$  (b) Non-helical magnetic energy in units of the initial non-helical magnetic energy; (c)  Total magnetic energy in units of the initial magnetic energy.  (d) Dimensionless small scale magnetic helicity $h_2$ also normalized to $h_c$.}
\label{fig1}
\end{figure}

\newpage
\begin{figure}
\centering
\mbox{\subfigure[]{\includegraphics[width=2.5in]{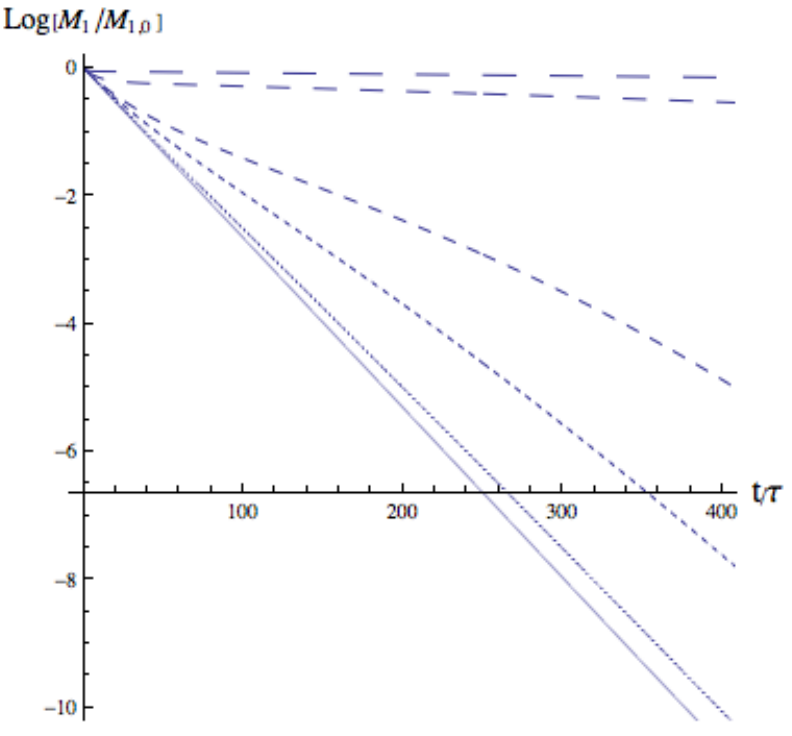}}
}
\caption{Solutions to Eqs. (\ref{9}), (\ref{10}), and (\ref{11})  for total large scale magnetic energy
using a fixed $f_{1,0}=0.999$, $k_1=1$ and $k_2=5$ with $R_M=800$ 
for  various  initial magnetic energies $M_{1,0}$.  
The third curve from the top corresponds to our analytically derived critical value  
$M_{1,0}=M_{1,c}=k_1 h_c/k_2=(k_1/k_2)^2$.  
From top to  bottom the curves  correspond to  
${M_{1,0}\over (k_1/k_2)^2} =20, 5,1,0.5,0.1, 0.01$, respectively.
}
\label{fig2}
\end{figure}

\newpage
\begin{figure}
\centering
\mbox{\subfigure[]{\includegraphics[width=2.5in]{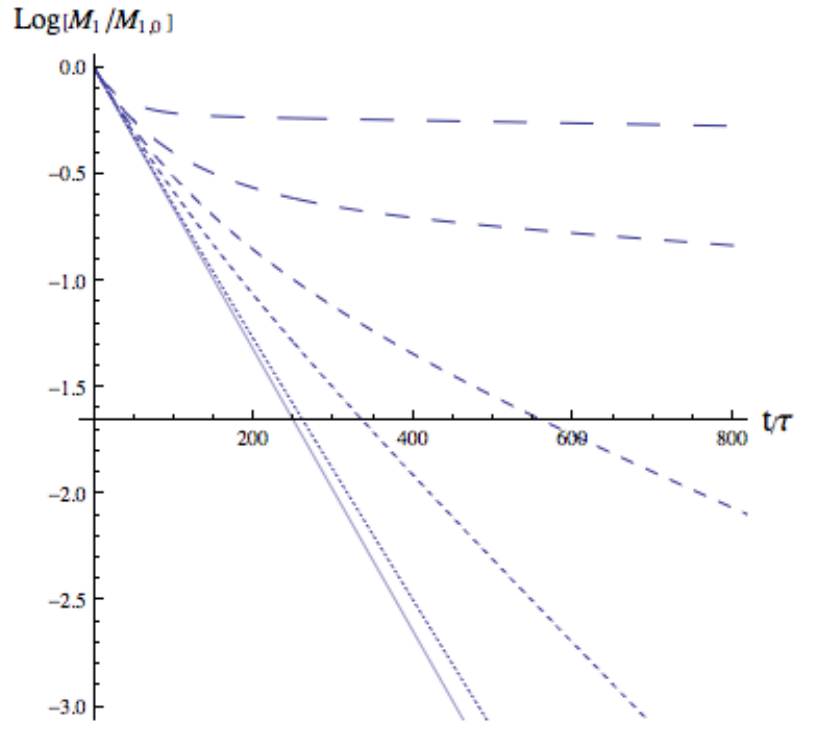}}\quad
\subfigure[]{\includegraphics[width=2.5in]{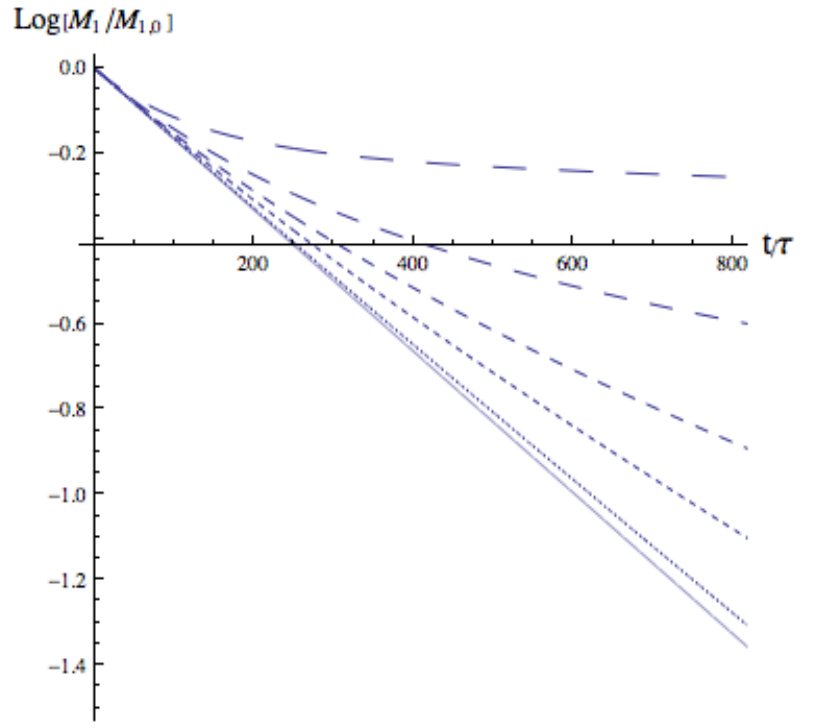}}}
\caption{ From top to bottom, the curves correspond to
 ${M_{1,0}\over (k_1/k_2)^2 }=5,2,1,0.5,0.1, 0.01$ respectively.
 for $k_2=10$ (left panel), $k_2=20$ (right panel), and
with $R_M=8000$ for both panels. }
\label{fig3}
\end{figure}

\bibliographystyle{mn2e}
\bibliography{diffusionlargescale10}
\end{document}